\documentclass[reprint,aps,prl,superscriptaddress,footinbib,longbibliography]{revtex4-1}
\usepackage{graphicx}
\usepackage{amsmath}
\usepackage{color}
\usepackage[colorlinks, linkcolor= blue, citecolor = blue, urlcolor=blue]{hyperref}
\usepackage[normalem]{ulem}

\begin{document}
\title{Photonic Topological Transitions and Epsilon-Near-Zero Surface Plasmons in Type-II Dirac Semimetal NiTe$_2$}
\author{Carlo Rizza} \email{email: carlo.rizza@univaq.it}
\affiliation{Department of Physical and Chemical Sciences, University of L'Aquila, Via Vetoio 1, I-67100 L'Aquila, Italy}
\author{Debasis Dutta}
\affiliation{Department of Physics, Indian Institute of Technology Kanpur, Kanpur-208016, India} 
\author{Barun Ghosh}
\affiliation{Department of Physics, Indian Institute of Technology Kanpur, Kanpur-208016, India}
\affiliation{Department of Physics, Northeastern University, Boston, Massachusetts 02115, United States of America}

\author{Francesca Alessandro}
\affiliation{Department of Physics, University of Calabria, Via ponte Bucci, I-83036, Rende, Italy}

\author{Chia-Nung Kuo}
\affiliation{Department of Physics, National Cheng Kung University, 1 Ta-Hsueh Road 70101 Tainan, Taiwan}

\author{Chin Shan Lue}
\affiliation{Department of Physics, National Cheng Kung University, 1 Ta-Hsueh Road 70101 Tainan, Taiwan}

\author{Lorenzo S. Caputi}
\affiliation{Department of Physics, University of Calabria, Via ponte Bucci, I-83036, Rende, Italy} 

\author{Arun  Bansil}
\affiliation{Department of Physics, Northeastern University, Boston, Massachusetts 02115, United States of America}

\author{Amit  Agarwal}
\affiliation{Department of Physics, Indian Institute of Technology Kanpur, Kanpur-208016, India}

\author{Antonio Politano}  \email{email: antonio.politano@univaq.it}
\affiliation{Department of Physical and Chemical Sciences, University of L'Aquila, Via Vetoio 1, I-67100 L'Aquila, Italy}
\affiliation{CNR-IMM Istituto per la Microelettronica e Microsistemi, VIII strada 5, I-95121 Catania, Italy} 

\author{Anna Cupolillo}
\affiliation{Department of Physics, University of Calabria, Via ponte Bucci, I-83036, Rende, Italy}

\begin{abstract}
Compared to artificial metamaterials, where nano-fabrication complexities and finite-size inclusions can hamper the desired electromagnetic response, several natural materials like van der Waals crystals hold great promise for designing efficient nanophotonic devices in the optical range.  Here, we investigate the unusual optical response of NiTe$_2$, a van der Waals crystal and a type-II Dirac semimetal hosting Lorentz-violating Dirac fermions. By {\it ab~initio~} density functional theory modeling, we show that NiTe$_2$ harbors multiple topological photonic regimes for evanescent waves (such as surface plasmons) across the near-infrared and optical range. By electron energy-loss experiments, we identify surface plasmon resonances near the photonic topological transition points at the epsilon-near-zero (ENZ) frequencies $\approx 0.79$, $1.64$, and  $2.22$  eV. Driven by the extreme crystal anisotropy and the presence of Lorentz-violating Dirac fermions, the experimental evidence of ENZ surface plasmon resonances confirm the non-trivial photonic and electronic topology of NiTe$_2$. Our study paves the way for realizing devices for light manipulation at the deep-subwavelength scales based on electronic and photonic topological physics for nanophotonics, optoelectronics, imaging, and biosensing applications. 
\end{abstract}


\maketitle
Artificial metamaterial structures offer a unique platform for achieving novel light-matter interaction regimes \cite{Cai}. For example, epsilon-near-zero (ENZ) regime \cite{Maas,Alam,Liberal}, extreme anisotropic responses \cite{Elser,Poddubny,Cop} are achieved in metamaterials in the framework of homogenization theories, where the metamaterial response is merely described by an effective dielectric permittivity tensor. However, the homogeneity assumption breaks down when the electromagnetic (EM) wavelength becomes comparable to or smaller than the size of metamaterial inclusions, hampering the 
the realization of the desired effective EM property, especially in the nanophotonic regime. 
\begin{figure}[t]
\centering
\includegraphics[width=.95\linewidth]{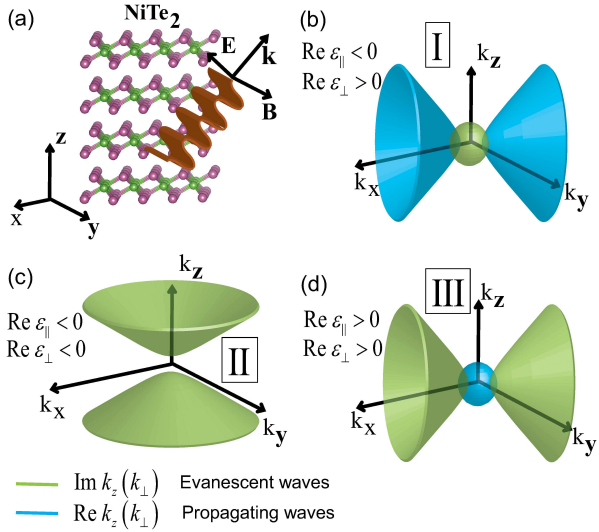}
\caption{(a) Schematic of an EM wave propagating in NiTe$_2$ sample, supporting different topological propagation/evanescent regimes in the optical range. (b)-(d) The three different isofrequency  surfaces (type I, II, and III) for propagating (cyan surfaces) and evanescent (green surfaces) waves in NiTe$_2$, achieved by varying the frequency in the near-infrared/optical range.
\label{Fig1}}
\label{Figure1}
\end{figure}

On the other hand, some natural materials have been reported to exhibit EM `meta'-responses
overcoming some limits of artificial composite metamaterials. Natural van der Waals crystals, stacked structures of atomically thin layers, hold great promise as an alternative platform to the anisotropic and even hyperbolic metamaterials \cite{HYPE,Choe,Gjerding,Boo,H_Gao,Lee} exhibiting peculiar optical properties such as negative refraction, strong enhancement of spontaneous emission, and  spatial filtering, among others \cite{Rho,LuLu,Sreekanth,Sim,Rizza,Vas}.

Remarkably, a class of van der Waals crystals supports topologically protected Weyl and Dirac semimetals in which the low-energy electronic excitations are described by Weyl and Dirac equations and, they also offer novel opportunities for photonics applications \cite{Hu,Sonowal,Jalali,Halter}. Recently, the transition-metal dichalcogenides (TMDs) TMX$_2$ (TM=Pd, Pt; X=Se, Te) have been demonstrated to host Lorentz-violating type-II Dirac fermions \cite{Clark1, Clark2}, where the Dirac cone is highly tilted in the momentum space \cite{Chang}. However, the Dirac point in most of these TMDs is located well below the Fermi level \cite{Clark1, Clark2,Yan,Huang}, so that the impact of the topological quasiparticles on the Fermi-surface-dependent physical properties gets smeared out \cite{Xuquadratic}. In contrast, the Dirac point in NiTe$_2$ lies close to the Fermi level, which has stimulated considerable interest \cite{Ferreira,libozhang,Qi_1,Nappini,Ghosh_1,Wang_1,Liu_1,Zhao_1,Li_1}. For example, NiTe$_2$ has been exploited for engineering long-wavelength photodetectors such as large-area ultrafast Terahertz imaging systems \cite{libozhang}, and microwave receivers \cite{Nappini}.


In this Letter, we unveil the unique EM response of NiTe$_2$ in the near-infrared/optical range arising from its extreme anisotropy driven by the presence of Lorentz-violating Dirac fermions. In particular, we demonstrate that NiTe$_2$ supports three photonic regimes with distinct topologies for evanescent waves (which have a rapidly decaying field amplitude in some spatial directions, such as surface plasmons) across the near-infrared/visible range. By tuning the optical frequency, NiTe$_2$ can exhibit closed elliptical and two different kinds of open hyperbolic dispersion, as shown in Fig.~\ref{Fig1}. Exploiting electron energy loss spectroscopy and $ab~ initio$ density functional theory, we identify multiple ENZ surface plasmons, exciting near photonic topological transitions (at about $0.79$, $1.64$, and $2.22$ eV), whose presence confirms the unusual topological photonic behavior of NiTe$_2$. It is worth noting that surface plasmons display several significant features (viz., a resonant nature, strong light enhancement, and a high degree of confinement) and have been exploited in an enormous variety of applications  \cite{wave,lens,ses}. Finally, we show that the type-II Dirac bands are strongly involved in the origin of the ENZ surface plasmons. Thus, these findings establish NiTe$_2$ as a suitable candidate for applications where an efficient manipulation of evanescent waves is crucial, viz., in nanophotonics, optoelectronics, imaging, and biosensing. 

%
%

In order to highlight the different photonic regimes, we consider a $p-$polarized plane wave, $\exp{\left( i {\bf k_{\parallel}} \cdot {\bf r_{\parallel}}+i k_{\perp} z -i \omega t \right)}$ with ${\bf r}_{\parallel}=x \hat{\bf e}_x+ y \hat{\bf e}_y$, ${\bf k_{\parallel}}=k_x \hat{\bf e}_x+ k_y \hat{\bf e}_y$, and $k_{\perp}=k_z$ excited inside a NiTe$_2$ crystal. The dispersion relation for this EM mode is given by 
\begin{equation}
\label{disp}
\frac{k_{||}^2}{\varepsilon_{\perp}(\omega)}+\frac{k_{\perp}^2}{\varepsilon_{||}(\omega)}=\frac{\omega^2}{c^2}.
\end{equation}
Here, $\varepsilon_{\parallel/\perp}(\omega)$ is the frequency-dependent in-plane/out-of-plane component of the dielectric permittivity tensor and $c$ is the speed of light in vacuum.  The different regimes of the propagating ($k_{\perp}$ is real) and evanescent ($k_{\perp}$ is imaginary) EM wave can be deduced from the iso-frequency dispersion specified in Eq.~\eqref{disp}. Depending on whether ${\rm Re} \varepsilon_{\parallel/\perp}(\omega) >0~({\rm or} <0$), we can get different topology of the iso-frequency surface of the propagating or evanescent waves.  Specifically, we find that NiTe$_2$ can support multiple regimes for the evanescent waves rapidly decaying perpendicular to the $x-y$ plane, as shown in Fig.~\ref{Fig1}. By tuning light frequency across the near-infrared/visible range, the dynamics of evanescent waves is described by either elliptical (I) or hyperbolic (II, III) dispersion relation (green surfaces in Fig.~\ref{Fig1}). In contrast, the bulk propagating waves have only hyperbolic (I) and elliptic (III) dispersion (cyan surfaces in Fig.~\ref{Fig1}). The bulk propagating waves are completely forbidden in the frequency regime where both ${\rm Re}\varepsilon_{\parallel/\perp}(\omega)<0$. Together, these features make NiTe$_2$ an ideal platform for harnessing the light-matter interactions in the deep-subwavelength regime \cite{Krishnamoorthy}. 

\begin{figure}[t]
\centering
\includegraphics[width=1.02\linewidth]{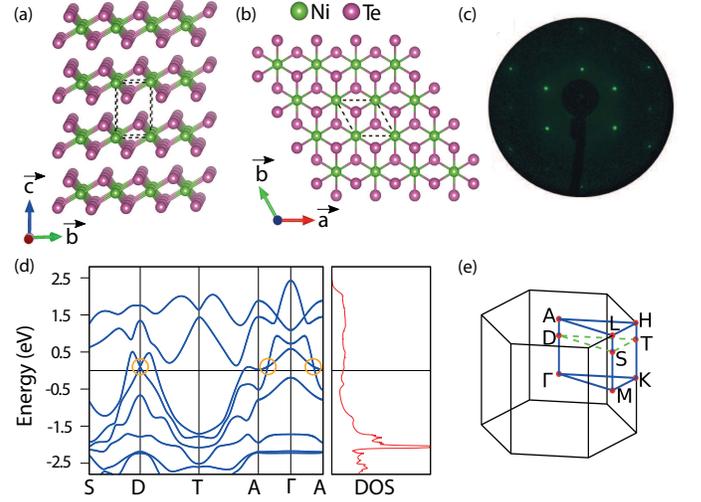}
\caption{(a) Side and (b) top views of the CdI$_2$ type trigonal atomic structure of NiTe$_2$. (c) LEED pattern with a primary electron beam energy of 84 eV. The hexagonal surface symmetry can be clearly seen. (d) Bulk band structure of NiTe$_2$. Orange circles highlight the pair of type-II Dirac points along the $\Gamma-A$ axis. 
(e) The bulk Brillouin zone of NiTe$_2$, with high symmetry points marked. `D' marked the type-II Dirac point.
\label{Fig2}}
\end{figure}

Bulk NiTe$_2$ crystallizes in the CdI$_2$ -type trigonal (1T) structure with space group $P{\bar 3}m1$ (No.~164), as illustrated in Fig.~\ref{Fig2}(a) and (b). Details of the growth and characterization of single crystals with X-ray diffraction and Raman spectroscopy are reported in \cite{SM}. The as-cleaved (001) surface displays a hexagonal surface symmetry, as evidenced by low-energy electron diffraction [LEED, shown in Fig.~\ref{Fig2}(c)]. The calculated band structure (Fig.~\ref{Fig2}d) displays the presence of a pair of type-II Dirac cones in the vicinity of the Fermi energy, along the $\Gamma-A$ axis. Angle-resolved photoemission spectroscopy experiments have confirmed the Dirac points to be located within $20$ meV above the Fermi energy, in addition to the presence of spin-momentum-locked surface states \cite{Ghosh_1}. The combination of hybridization of the Te-p orbital manifolds, crystal field splitting, and spin-orbit coupling (SOC) gives rise to multiple other topological features in NiTe$_2$ over a wide energy range.

We evaluate the frequency-dependent dynamical dielectric functions in the long-wavelength limit $\varepsilon_{\perp/\parallel}(\omega, {\bf q} \to 0)$, within the random phase approximation (RPA) \cite{Agarwal,politano,Sadhukhan}. Specifically, starting from the generalized gradient approximation-based density functional theory, including the SOC, $\varepsilon_{\perp/\parallel}(\omega)$ are calculated from an accurate Wannier tight-binding model of NiTe$_2$ (see Supplementary Materials for details \cite{SM}).

\begin{figure}
\centering
\includegraphics[width=0.5\textwidth]{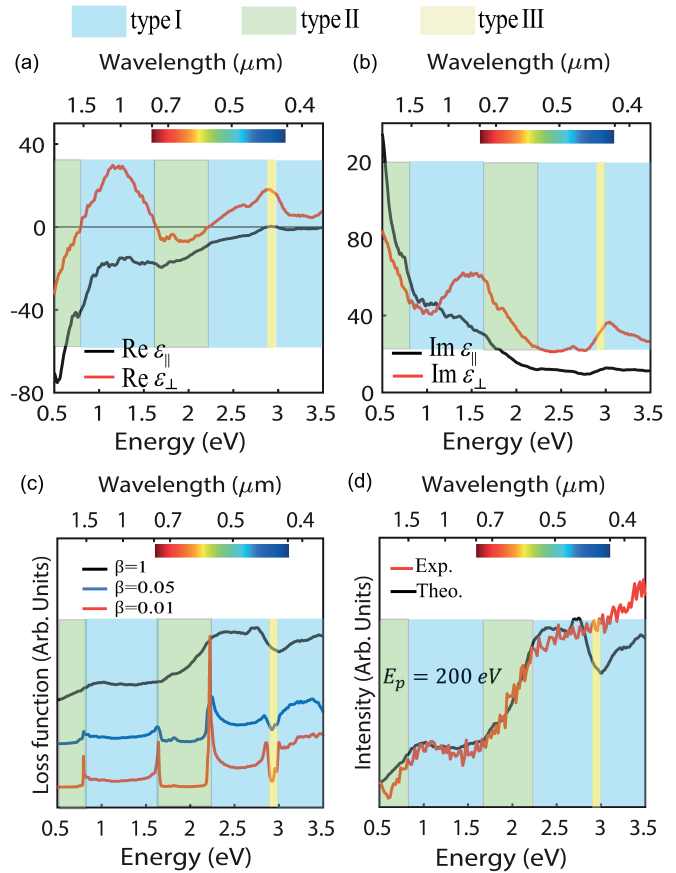}
\caption{(a) Real and (b) imaginary part of the in-plane $\varepsilon_{\parallel}$ and out-of-plane $\varepsilon_{\perp}$ dielectric functions of NiTe$_2$.  (c) Theoretical energy loss functions, in which the impact of the EM losses on ${\rm Im}\varepsilon_{\parallel}(\omega)$ and ${\rm Im}\varepsilon_{\perp}(\omega)$ is assessed through the inclusion of a factor $\beta$ equal to $1,0.05,0.01$ (black,blue,red solid lines, respectively), useful in order to clearly disclose the topological transitions in NiTe$_2$ in the optical range. (d) Comparison between the experimental EELS spectrum, at a primary electron beam energy of $E_p=200$ eV, (red solid line) and the theoretical energy loss function with $\beta=1$ (black solid line). The light blue, green and yellow areas correspond to the three regimes I,II, and III reported in Fig.1 (b), (c) and (d), respectively. The color bar in all the panels marks the optical spectral range. 
\label{Fig3}}
\end{figure}

The unconventional EM response of NiTe$_2$ is unveiled by the behavior of its frequency dependent dielectric functions, showing three distinct topological regimes (viz., I, II, and III as reported in Fig.~\ref{Fig1}) marked by green, blue and yellow region in Fig.~\ref{Fig3}, respectively. In the blue region of Fig.~\ref{Fig3}(a) and (b) [$0.79$ $<\hbar \omega<$   $1.64$ eV and $2.22$  $<\hbar \omega<$   $2.88$ eV, and $\hbar \omega>$   $2.98$ eV], we have ${\rm Re}\varepsilon_{\perp}(\omega)>0$ and ${\rm Re} \varepsilon_{||}(\omega) < 0$. Consequently, in these regions, the evanescent (propagating) waves experience an elliptic (hyperbolic) dispersion. In the regions depicted in green in Fig.~\ref{Fig3}, bounded by $\hbar \omega < 0.79 $ eV, and $1.64$ $<\hbar \omega<$   $2.22$ eV, we have ${\rm Re} \varepsilon_{\perp}(\omega)<0$ and ${\rm Re}\varepsilon_{||}(\omega)<0$. Thus, in these regions, there are no propagating plane waves, though evanescent waves such as surface plasmons can be excited. In the relatively narrow spectral range  $2.22$ $<\hbar \omega<$ $2.98$ eV, marked by yellow in Fig.~\ref{Fig3}, NiTe$_2$ acts as an anisotropic positive dielectric medium, with ${\rm Re} \varepsilon_{\perp/\parallel}(\omega) >0$. In this regime, the evanescent (propagating) waves exhibit the standard hyperbolic (elliptic) dispersion. 

To probe the evanescent surface modes of NiTe$_2$, we use electron loss energy spectroscopy (EELS), which is an ideal tool for detecting surface plasmons \cite{politano, rocca1, rocca2}. Details of the experimental methods are presented in Supplementary Materials \cite{SM}. EELS can probe surface plasmons when the electron penetration is negligible and the bulk modes are not excited, as for the case of low-energy impinging electrons (E$_p \leq$ 300 eV) in the reflection geometry. In a uniaxial crystal probed in the reflection geometry, the experimental EELS spectrum is theoretically reproduced by the surface loss function, {$P_{\rm surf}(\omega)$} \cite{Lam}, 
\begin{equation}
\label{PP}
P_{\rm surf}(\omega)=\frac{e^2}{4 \pi \varepsilon_0 \hbar v_{\perp}} \frac{1}{\omega} {\rm Im} \left[\frac{\xi(\omega)-1}{\xi(\omega)+1} \right].
\end{equation}
Here, $v_{\perp}$ is the component of the electron velocity along the optical $z$-axis,  
and $\xi(\omega)=\sqrt{\varepsilon_{||}(\omega)\varepsilon_{\perp}(\omega)}$ (with ${\rm Im} \xi >0$) is the geometrical average of the dielectric tensor components ($\hbar$ is the reduced Planck's constant) \cite{Lam}. The energy loss spectrum of Eq.~\eqref{PP} is peaked at a frequency for which ${\rm Re}\xi(\omega)=-1$, corresponding to the excitation of a surface plasmon (see \cite{SM} for further details). We evaluate the surface loss function of Eq.~\eqref{PP} for NiTe$_2$ by using the dielectric functions calculated from first principles reported in Fig.~\ref{Fig3}(a)-(b). 

To reveal the connection between the different topological EM propagation regimes of NiTe$_2$ and the experimental EELS spectrum, we investigate the case, where the imaginary parts of the dielectric components are diminished by a factor $\beta$, viz., $\varepsilon_{\perp/\parallel}(\omega)={\rm Re} \varepsilon_{\perp/\parallel}(\omega)+i \beta {\rm Im} \varepsilon_{\perp/\parallel}(\omega)$. We plot $P_{\rm surf}(\omega)$ in Fig.~\ref{Fig3}(c), where $\beta=1,0.05,0.01$ (black, blue and red line, respectively). For small EM losses ($\beta=0.01$), several peaks, coinciding with the topological transitions for EM propagation, clearly emerge in the spectrum. In particular, we find three peaks in the surface loss function at $0.79$, $1.64$, and $2.22$ eV, all of which lie near the boundary of the region where NiTe$_2$ behaves as a negative dielectric medium (green shaded regions with both ${\rm Re}\varepsilon_{\parallel/\perp}(\omega) <0$). All three surface plasmon peaks arise from the anisotropic dielectric function of NiTe$_2$, which in turn originates from the structural anisotropy and the resulting anisotropy in the electronic dispersion. Corresponding to high-momentum surface plasmon polaritons \cite{Warm}, the condition for the existence of a surface plasmon is $\xi(\omega)=-1$.  This is exactly satisfied in the ideal case where the imaginary parts of dielectric functions are negligible for $ \varepsilon_{\perp}(\omega)=1/\varepsilon_{\parallel}(\omega)$. As a consequence, in NiTe$_2$ where ${\rm Re}\varepsilon_{\parallel}(\omega) \ll -1$, the surface plasmon resonances in NiTe$_2$ are excited around the ENZ crossing points ${\rm Re} \varepsilon_{\perp}(\omega) \simeq 0$. Thus, the plasmonic resonances identify the ENZ conditions along with the location of the photonic topological transitions associated with the dispersion of evanescent waves as reported in Fig.~\ref{Fig1}(b) and (c). In addition to the three plasmon peaks arising in vicinity of ${\rm Re}\varepsilon_{\perp}(\omega) \approx 0$, there are two additional peaks at the boundary of the yellow region (at about $2.88$  and $2.98$ eV) arising from the ENZ conditions ${\rm Re}\varepsilon_{\parallel}(\omega)\simeq 0$ (and $\xi \simeq 0$). These peaks correspond to the topological photonic transitions between the EM dispersion regimes reported in Fig.~\ref{Fig1}(b) and (d).

\begin{figure}[t!]
\centering
\includegraphics[width=\linewidth]{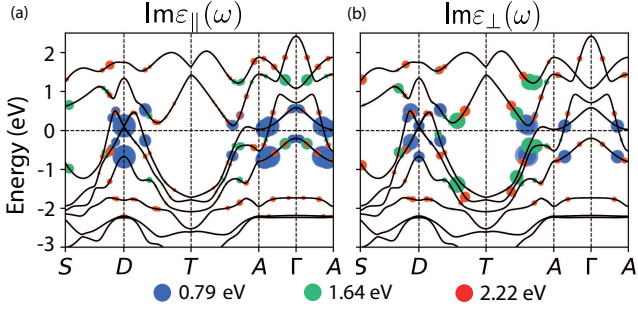}	
\caption{Weights of different bands participating in the vertical interband transitions giving rise to the imaginary part of the (a) in-plane, ${\rm Im} \varepsilon_{\parallel}(\omega)$, and (b) out of plane, ${\rm Im} \varepsilon_{\perp}(\omega)$, dielectric functions evaluated at the surface plasmon frequencies. Different colors refer to the different plasmon energies, viz., the $0.79,1.64,2.22$ eV surface plasmon modes are in blue, green, and red filled circles, where the circle size is proportional to the contribution.
\label{Fig4}}
\end{figure}

The experimental EELS spectrum is shown in Fig.~\ref{Fig3}(d), along with the theoretical loss function (for $\beta = 1$). Our experimental results are in good agreement with the theoretical predictions, validating the occurrence of exotic topological photonic transitions in NiTe$_2$. The measured spectra show two `flat' regions (with reduced curve slope) approximately corresponding to the two hyperbolic zones $0.79$ $<\hbar \omega<$ $1.64$ eV and $2.22$ $<\hbar \omega<$   $2.88$ eV. The first flat region ($0.79$ $<\hbar \omega<$ $1.64$ eV) arises from the superposition of the two surface plasmons exhibiting a relatively large resonance width, due to large EM losses (${\rm Im}\varepsilon_{\parallel/\perp}(\omega)\gg 0$). The second flat region ($2.22$ $<\hbar \omega<$ $2.88$ eV) arises from the superposition of the resonant surface plasmon at $2.22$ eV and the peak in the loss function related to the unusual ENZ response (${\rm Re}\varepsilon_\parallel(\omega) = 0$) of NiTe$_2$.  Note that surface plasmons are high-momentum evanescent waves (viz., $k_{\parallel}\gg \omega/c$) \cite{Warm,Zay} and, generally, they are difficult to be excited in the standard metamaterials, since the effective medium theory does not hold when the EM momentum approaches the inverse size of the metamaterial inclusions. Thus, the desired metamaterial response can be completely washed out by the spatial scales associated with the metamaterial inclusions. In contrast, here, the presence of surface plasmons confirms that NiTe$_2$ preserves the EM response even in the deep subwavelength regime.    

Considering the possible usage of the observed plasmonic modes for technological applications, we validated the robustness of the observed plasmonic modes in NiTe$_2$. Furthermore, We conducted a systematic investigation of the dependence of the EELS excitation spectrum on the aging phenomena in the sample kept in air. The EELS spectrum does not display noticeable changes upon exposure to air, congruently with ambient stability inferred by AFM experiments (see \cite{SM} for additional details).

Finally, to physically grasp the origin of all the three surface plasmon modes, we focus on bulk dielectric function and its connection to the electronic bands. 
The imaginary part of the band and momentum resolved dielectric function (in the long wavelength limit) can be expressed as ${\rm Im}\varepsilon_{\perp/\parallel}(\omega) = \sum_{{\bf q}^{'}, m}{\rm Im}\varepsilon^{(m)}_{\perp/\parallel}(\omega,{\bf q}=0,{\bf q}^{'})$, with $m$ being the band index, ${\bf q}^{'}$ representing a momentum point in the Brillouin zone (BZ).  Here, we have defined the momentum and band resolved dielectric function,  
\begin{eqnarray}
& & {\rm Im}\varepsilon^{(m)}_{\perp/\parallel}(\omega,{\bf q}=0,{\bf q}^{'}) =  
\frac{\pi e^2}{\varepsilon_0\Omega_{\rm cell}}\sum_{n\ne m} (f_{m,{\bf q}^{'}} -f_{n,{\bf q}^{'}}) \\ 
& & \times \frac{ \langle{m,{\bf q}^{'}}|{\hat v}_{\perp/\parallel}|n,{\bf q}^{'}\rangle\langle{n,{\bf q}^{'}}|{\hat v}_{\perp/\parallel} |m,{\bf q}^{'}\rangle }{\Delta E_{m,n,{\bf q}^{'}}^2}
\delta \left( \hbar \omega+\Delta E_{m,n,{\bf q}^{'}} \right). \nonumber 
\label{Eq_3} 
\end{eqnarray}
In Eq.~\eqref{Eq_3}, $f_{n,{\bf q}^{'}}$ is the Fermi function, $E_{n,{\bf q}^{'}}$ represents the band dispersion of the $n$th band, $\Delta E_{m,n,{\bf q}^{'}}=E_{m,{\bf q}^{'}}-E_{n,{\bf q}^{'}}$, $\Omega_{\rm cell}$ denotes unit cell volume, and $\hat{v}_{i} = \partial_{q'_{i}} {\cal H}$ are the perpendicular/parallel ($i=\perp/\parallel$) components of the velocity operator (${\cal H}$ being the Hamiltonian of the system). Equation \eqref{Eq_3} captures the `strength' of the vertical (${\bf q}=0$) interband electronic correlation ($E_{m,{\bf q}^{'}} \to E_{n,{\bf q}^{'}}$) at the ${\bf q}^{'}$ point in the BZ \cite{SM}. Figure~\ref{Fig4} shows the relative strength of ${\rm Im}\varepsilon^{(m)}_{\perp/\parallel}(\omega,{\bf q}=0,{\bf q}^{'})$, projected along the high symmetry directions in the BZ, for the three surface plasmon frequencies. In addition to contributions from higher conduction bands, the $2.22$ eV surface plasmon (red circles in Fig.~\ref{Fig4}) has spectral contributions arising from interband correlations involving the lowest conduction band hosting the Dirac fermions. The $1.64$ eV surface plasmon (green circles in Fig.~\ref{Fig4}) predominantly involves electronic correlations from the lower and the highest valance band ,including the Dirac bands, to higher conduction bands. The $0.79$ eV surface plasmon (blue circles in Fig.~\ref{Fig4}) has significant contribution arising from the interband electronic correlations involving the tilted type-II Dirac bands of NiTe$_2$, in the vicinity of the Fermi energy.

In conclusion, we demonstrate that NiTe$_2$ exhibits an unusual EM response in the near-infrared and visible frequency regime, driven by its huge intrinsic EM anisotropy and the presence of Lorentz-violating Dirac fermions. The isofrequency surfaces of evanescent EM waves in NiTe$_2$ display several different photonic topologies, including one elliptical and two distinct hyperbolic dispersions. Our EELS experiments probe these photonic features as resonance peaks associated with the ENZ surface plasmons. Our study establishes NiTe$_2$ and related materials can behave as hyperbolic, ENZ, and extremely anisotropic photonic materials with a `perfect' homogeneous EM response even for highly confined surface waves. Overcoming some limitations of artificial composite metamaterials, topological semimetals, such as  NiTe$_2$, offer a novel pathway for engineering optical properties in the nanophotonic regime.

\section*{ACKNOWLEDGEMENT}
B.G, D.D, and A.A thank the Science and Engineering Research Board and the Department of Science and Technology of the government of India for financial support. We thank the computer centre at IIT Kanpur, India for providing the high-performance computing facilities. The work at Northeastern University was supported by the Air Force Office of Scientific Research under award number FA9550-20-1-0322, and it benefited from the computational resources of Northeastern University's Advanced Scientific Computation Center (ASCC) and the Discovery Cluster.


\onecolumngrid
\begin{center}
  \textbf{Supplementary Material}\\[.2cm]
\end{center}

\setcounter{equation}{0}
\setcounter{figure}{0}
\setcounter{table}{0}
\setcounter{page}{1}
\renewcommand{\theequation}{S\arabic{equation}}
\renewcommand{\bibnumfmt}[1]{[S#1]}
\renewcommand{\citenumfont}[1]{S#1}

\newcommand{\bkt} [1] {\langle #1 \rangle}

\newcommand{\dps}{\displaystyle}
\newcommand{\para}{\parallel}
\newcommand{\om}{\iffalse}
\newcommand{\pd}[2]{\frac{\partial #1}{\partial #2}}

\renewcommand{\thepage}{S\arabic{page}}  
\renewcommand{\thesection}{S\arabic{section}}     
\renewcommand{\thetable}{S\arabic{table}}   
\renewcommand{\thefigure}{S\arabic{figure}}

\section{Details of the growth and experimental characterization of a single crystal}

\subsection*{Single-crystal growth}
Single crystals of NiTe$_2$ were grown using the Te flux method. We used various molar ratios of Ni to Te, ranging from $1 : 3$ to $1 : 30$, finding an optimum ratio of $1 : 8$. The mixtures of high-purity Ni powder ($99.99\%$) and Te ingots ($99.9999\%$) were sealed under vacuum in a quartz tube with a flat bottom. The quartz ampule was heated to $1050$ $^\circ{C}$ over $10$ hours, held at a constant temperature for $10$ hours, and then slowly cooled to $600$ $^\circ{C}$ at a rate of $3$ $^\circ{C}$/h. It was finally annealed at this temperature for $100$ hours to improve the quality of the crystals. The remaining Te flux was removed by centrifuging above $550$ $^\circ{C}$. Several shiny plate-like single crystals with typical sizes of $8 \times 8 \times 1 $ mm$^3$ were harvested, which could be easily cleaved with adhesive tape. The flat surface of the crystal corresponds to the $ (001)$ plane.

\subsection*{XRD}
The crystal structure and phase purity of the as-grown crystals were identified by X-ray diffraction (XRD) (Bruker D2 PHASER) and Laue diffraction (Photonic Science) at room temperature. Figure S1 (a) shows the powder XRD pattern of NiTe$_2$ samples, whose photograph is shown in the inset. Figure S1 (b) displays the single-crystalline XRD pattern from the (00l) planes of NiTe$_2$. The inset represents the Laue diffraction taken along the (0001) direction. 
\begin{figure}
\centering
\includegraphics[width=0.5\linewidth]{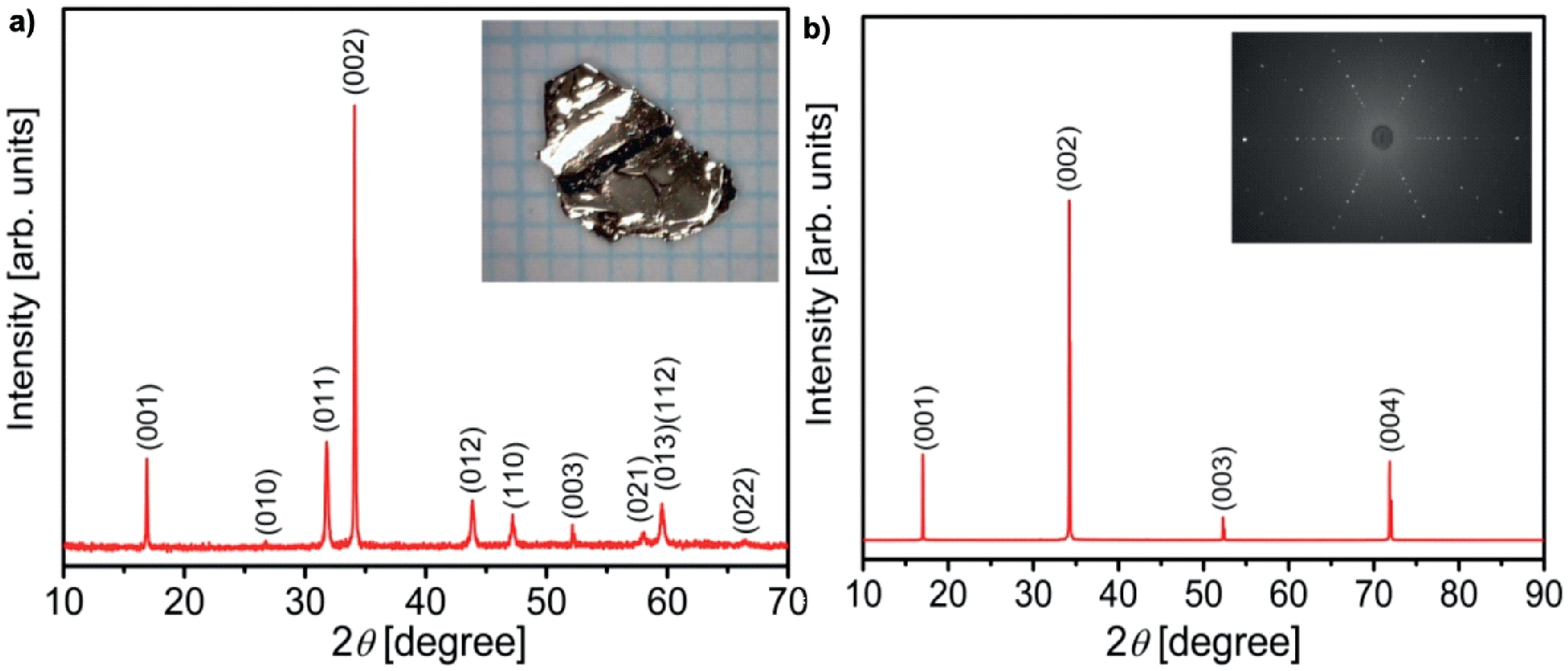}	
\caption{(a) XRD pattern of NiTe$_2$, taken with $Cu K_{\alpha}$. The inset shows the photo of grown single-crystal samples. (b) Single-crystalline XRD pattern of NiTe$_2$. The inset shows the Laue diffraction of NiTe$_2$ crystal.}
\end{figure}

\subsection*{Survey XPS spectrum of the as-cleaved sample}
Figure S2 shows the survey XPS spectrum of the as-cleaved sample, validating its surface cleanliness. No trace of contamination is observed. In particular, O-1s and C-1s are totally absent.
\begin{figure}
\centering
\includegraphics[width=0.5\linewidth]{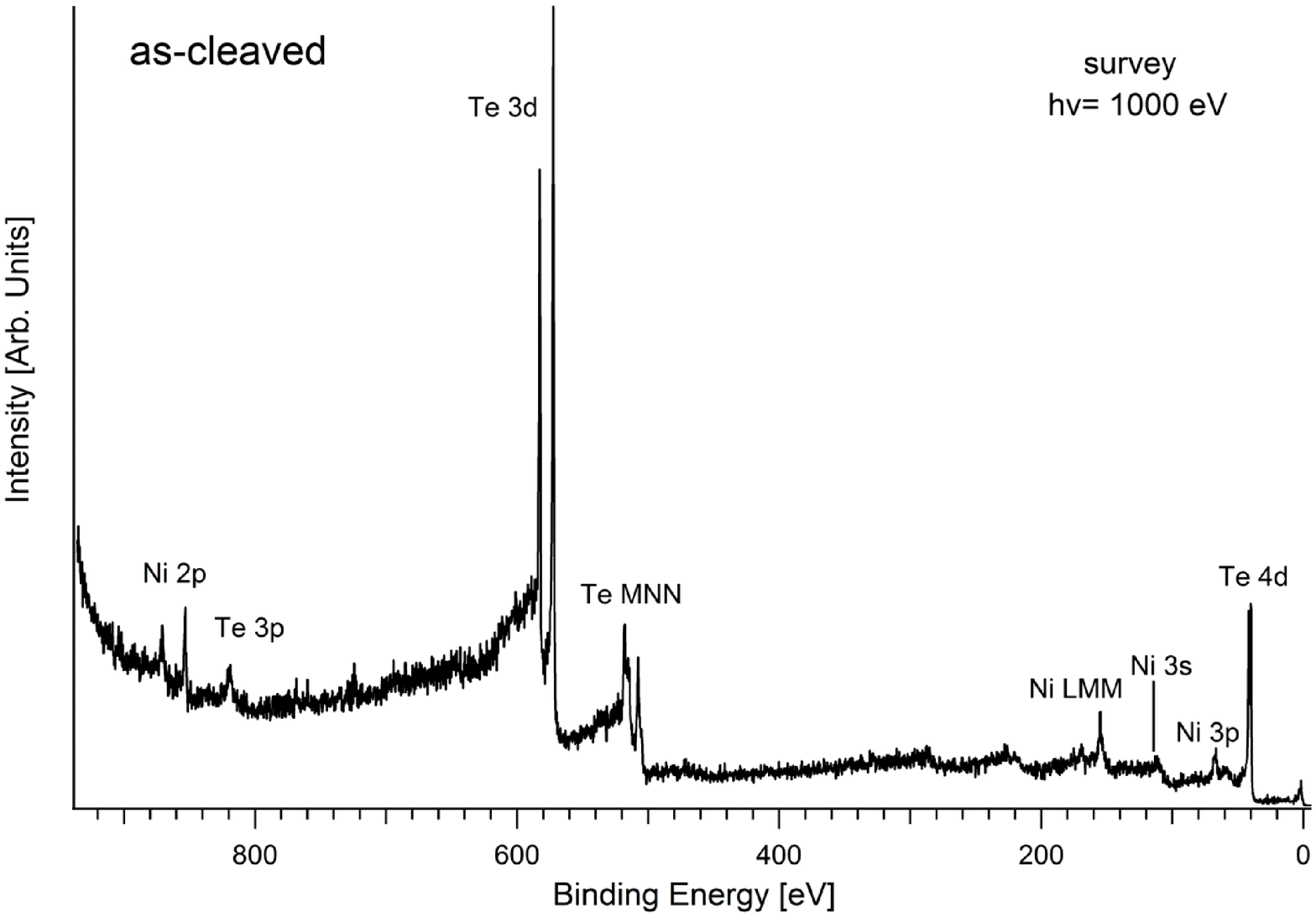}	
\caption{Survey XPS spectrum of as-cleaved sample of NiTe$_2$ measured with a photon energy of $1000$ eV}
\end{figure}

\subsection*{Raman spectroscopy}
Phonon dispersion calculations were performed within the \textit{frozen-phonon} method using the projector augmented wave (PAW) pseudopotentials and a plane basis set as implemented in VASP \cite{11,22,33}. The cut-off for the plane-wave basis is set to $500$ eV and a $\Gamma$-centered $12\times 12 \times 8$ $k$-grid was used for the BZ integration \cite{44}. The exchange-correlation part of the potential was treated within the local density approximation (LDA) \cite{55}.  All the structures were properly relaxed until the force on each atom becomes vanishingly small. Below, we have tabulated the phonon frequencies of bulk NiTe$_2$ in cm$^{-1}$ at the $\Gamma$-point, indicating their infrared or Raman activity based on our DFT calculations. The modes corresponding to $\sim 87$ and $\sim 150$ cm$^{-1}$ are Raman active.

Thus, in NiTe$_2$ there are $9$ phonon modes, $3$ acoustic, and $6$ optical. Only $E_g$ and $A_{1g}$ are Raman active. Correspondingly, two Raman-active modes are observed experimentally (Fig. S3).

In Figure S4, we show a sketch of the displacements of the atoms, corresponding to the $A_{1g}$ and one of the $E_g$ phonon modes. Note that both $E_g$ modes involve the motion of Te atoms primarily. 
\begin{figure}[h]
\centering
\includegraphics[width=0.4\linewidth]{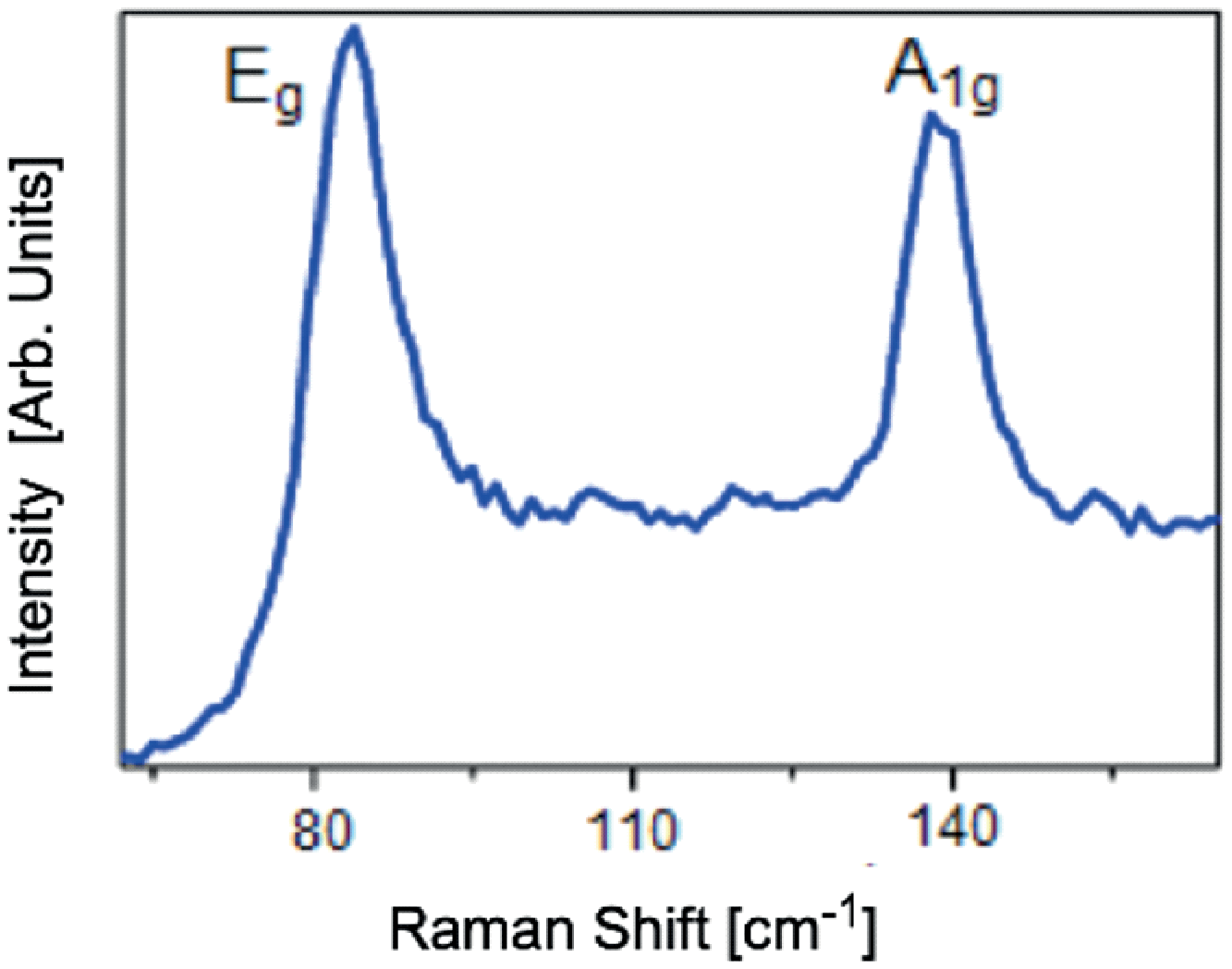}
\caption{Raman spectrum of NiTe$_2$.}
\end{figure}
\begin{table*}
\centering
\caption{Phonon frequencies of bulk NiTe$_2$ at the $\Gamma$-point.}
\begin{tabular}{|c |c |c |c|} 
\hline
Branch number & Frequency (cm$^{-1}$)  & Symmetry  & Infrared (I)/ Raman (R) activity \\
 
 1-2 & 0.00 & $E_u$ & - \\ 
 \hline
 3 & 0.00 & $A_{2u}$ & - \\
 \hline
 4-5 & 87.76 & $E_g$ & R \\
 \hline
 6 & 150.30 & $A_{1g}$ & R \\
 \hline
 7 & 211.58 & $A_{2u}$ & I \\ 
  \hline
 8-9 & 227.66 & $E_{u}$ & I \\ 
 \hline
\end{tabular}
\end{table*}
\begin{figure}
\centering
\includegraphics[width=0.8\linewidth]{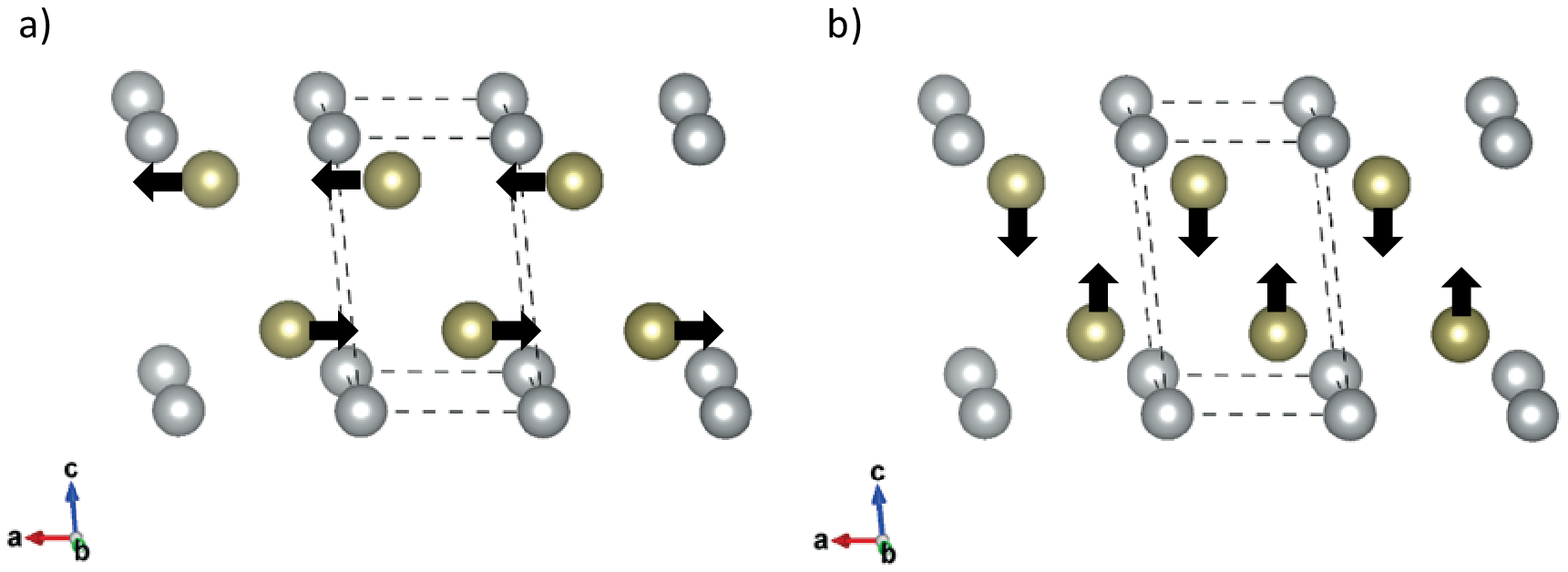}	
\caption{Schematic representation of (a) one of the $E_g$ modes and (b) the $A_{1g}$ phonon. Gray and golden spheres indicate Ni and Te atoms, respectively.}
\end{figure}
\section{Details of the electronic structure calculations}
To understand the electronic band structure of NiTe$_2$, we perform a first-principle calculation using Vienna $Ab~initio$ Simulation Package~(VASP)~\cite{11}. The projector augmented wave (PAW) pseudopotentials have been adopted within the generalized gradient approximation (GGA) scheme developed by Perdew-Burke-Ernzerhof (PBE)~\cite{22,Perdew1996}. We have used kinetic energy cutoff of $400$ eV for the plane-wave expansion and a $15\times{15}\times{10}$ Monkhrost-Pack $k-$grid for the Brillouin zone integration~\cite{mp_kgrid}. Spin-orbit coupling (SOC) has been considered self-consistently. 

\begin{figure}
	\includegraphics[width = 1\textwidth]{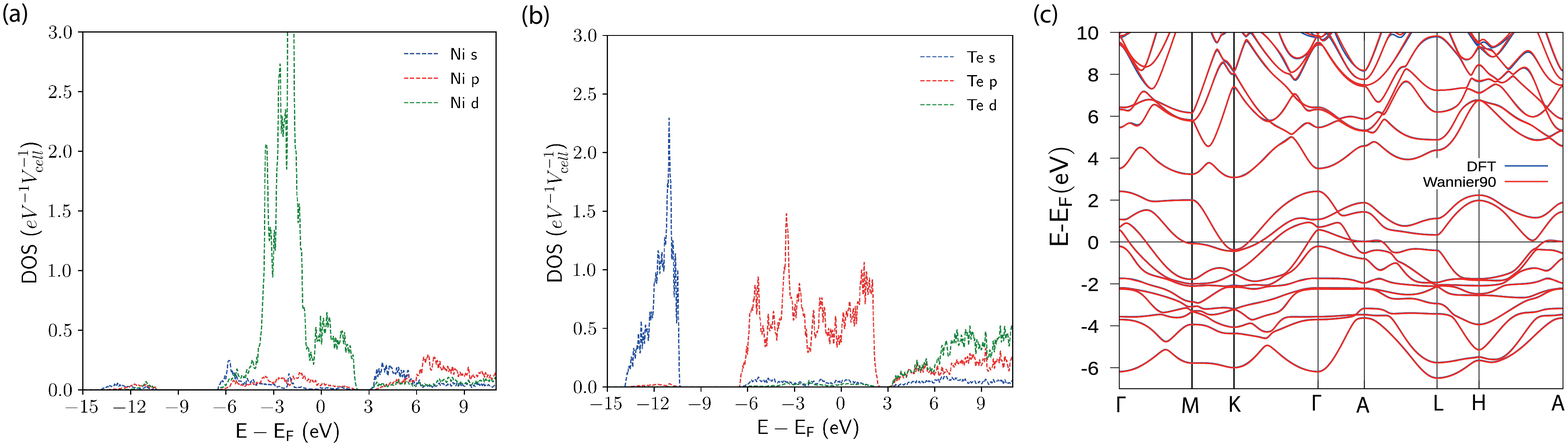}
	\caption{ (a), (b) Orbital resolved density-of-states (DOS) of NiTe$_2$, calculated using the PBE exchange correlation function in density functional theory. We find that the dominant contribution in the DOS comes from the 
	 $s$, $p$ and $d$ orbitals of Ni, and the $p$ and $d$ orbitals of Te. Thus, we have retained these orbitals to construct an accurate Wannier model over a large energy range, used for the dielectric tensor calculations. (c) The fit between the DFT band-structure and the Wannier TB hamiltonian over an energy window of [-7,10]eV is shown.}
	\label{fig_orbital_bs}
\end{figure}

In order to numerically evaluate the dielectric tensor, we construct a tight-binding (TB) model based on maximally localized Wannnier functions (MLWF)~\cite{Marzari2012}.  To make a good Wannier model valid in a wide range of energies for the DFT bands, we have included all the dominant orbitals present in the required energy range as Wannier orbitals. The orbital resolved density of states for the electronic bands of NiTe$_2$ over a wide energy range is shown in Fig.~\ref{fig_orbital_bs}. Based on this, we have included the $s$, $p$, and $d$ orbitals of Ni and $p$ and $d$ orbitals for Te atoms.
\section{Dielectric function calculation}\label{density_response_and_EELS}
We have calculated diagonal components of frequency and momentum dependent dynamical dielectric tensor within random phase approximation (RPA)~\cite{PhysRevB.73.045112,giuliani2005quantum}
\begin{equation}
\varepsilon_{ij}({{q},\omega})=\delta_{ij}\varepsilon_{ii}({{q},\omega}), ~~~{\rm where}~~~ \varepsilon_{ii}({{q},\omega}) \equiv \left[1-V({q})\Pi^{0}(q{\hat {\bf e}}_{i},\omega)\right].
\label{epsiloneq}
\end{equation}
Here, $i/j=x,~y,~z$, denote the three cartesian components, and $\hat{\bm e}_{i}$ are the unit vectors along these directions. Here $V(q)=e^2/{(\varepsilon_0 q^2)}$ is the Fourier transform of Coulomb potential in three-dimension, with $\varepsilon_0$  denoting the vacuum permittivity. In Eq.~\eqref{epsiloneq},  $\Pi^{0}({ {q}{\hat {\bm e}}_{i},\omega})$ represents the noninteracting density-density response function or polarization, calculated along different directions.  

To calculate the dielectric tensor, we first evaluate the dynamical form factor\cite{Thygesen2011} by using the energy eigenvalues and eigenstates of the Wannier TB Hamiltonian with SOC,  

\begin{eqnarray}
\label{form_fac}
S_{0}({ q\hat{\bm e}_{i}},\omega)= ~& & \frac{1}{\Omega_{\rm{cell}}}\sum_{{\bf q'},m,n}(f_{m,{\bf q'}}-f_{n,{\bf q'}+{ q\hat{\bm e}_{i}}})|{\langle{ {n,{\bf q'}+q\hat{\bm e}_{i}}|}{{m,\bf q'}}\rangle}|^2
\cdot\delta(\hbar\omega+E_{m,{\bf q'}} - E_{n,{\bf q'}+{q\hat{\bm e}_{i}}})~.
\end{eqnarray}
Here, $\hbar{{\bf q}}$ and $\hbar{\omega}$ denote the momentum and energy transfer to the quasiparticle and $E_{m,{\bf q'}}$ is the energy eigenvalue for the ${\bf q'}$ state of the $m$th band, $|m, {\bf q'}\rangle$ and $\Omega_{\rm cell}$ denotes the unit cell volume. Here the summation on ${\bf q'}$ runs over the first Brillouin zone (BZ) and the band occupancy is given by the Fermi distribution function, $f_{m,{\bf q'}}=\left[e^{{\beta(E_{m,{\bf q'}}-\mu)}}+1\right]^{-1}$, with $\mu$ denoting the chemical potential. For simplicity, we work in the zero temperature limit for which the Fermi Dirac distribution function reduces to a step function, $f_{m,{\bf q'}}=\theta(\mu-E_{m,{\bf q'}})$.
Using the dynamical from factor in Eq.~\eqref{form_fac}, the non-interacting density-density response function is obtained by using Kramers-Kronig transformation~\cite{Thygesen2011}, 
\begin{eqnarray}
\Pi^{0}({ q\hat{\bm e}_{i}},\omega)=\int_{0}^{\infty}d\omega^{\prime}S_{0}({ q\hat{\bm e}_{i}},\omega^{\prime})
\left[\frac{1}{\omega-\omega^{\prime}+i\eta}-\frac{1}{\omega+\omega^{\prime}+i\eta} \right].\nonumber
\end{eqnarray}
Here, $\eta$ is a small broadening parameter. We have used a relatively dense $k$-grid of $90\times{90}\times{60}$ in the calculation of the non-interacting response function.
Then, we compute the dynamical dielectric tensor using Eq.~\eqref{epsiloneq}. 
The direction of the momentum transfer $\bf{q}$ is chosen along specific high-symmetry directions in the Brillouin zone like $\Gamma-\rm{Z}$, $\Gamma-\rm{M}$ etc. Depending on the direction of ${ \bf{q}}$, we have computed different components of dielectric tensor as $\varepsilon_{xx }$, $\varepsilon_{yy}$, and $\varepsilon_{zz }$.  The rotational in-plane symmetry of the NiTe$_2$ crystal structure forces $\varepsilon_{{xx}}=\varepsilon_{{yy}}\equiv\varepsilon_{\rm{\parallel}}$. However, the out-of-plane component of the dielectric tensor, $\varepsilon_{zz} \equiv \varepsilon_{\perp}$ turns out to be different than the in-plane component due to the layered structure of NiTe$_2$. 

\section{EELS experiments}

The reflection EELS experiments were carried out at room temperature using an EELS apparatus with two $50$ mm hemispherical deflectors for both monochromator and rotating analyzer, mounted in an ultrahigh-vacuum chamber at the University of Calabria, Italy. All experiments were made in a base pressure $< 2 \cdot 10^{-10}$ mbar.  Spectra were acquired with an incident electron beam positioned at a fixed angle of 45$^\circ$ with respect to the surface normal. The kinetic energy of the incident electrons was 200 eV. For these experiments, scattered electrons were collected in specular reflection condition, with an angular acceptance of 2$^\circ$.  A medium energy resolution of $15$ meV was set to achieve the best signal-to-noise ratio.  

\section{Surface Plasmons in a uniaxial crystal}
Let us consider a system consist of a semi-infinite slab and a vacuum region, located in the half-space $z<0$ and $z>0$, respectively. The slab is a uniaxial crystal, where the $z$-axis is the crystal optical axis, and its dielectric permittivity tensor is $\varepsilon(\omega)=diag [\varepsilon_{\parallel}(\omega),\varepsilon_{\parallel}(\omega),\varepsilon_{\perp}(\omega)]$. Here, we discuss surface plasmons (SPs) supported by a uniaxial crystal (like NiTe$_2$). SPs belong to a subclass of SP polaritons, namely special $p$-polarized plane waves propagating along the vacuum-slab interface, whose amplitudes decay away from the interface\cite{Zay} $z=0$ .  Without loss of generality, we search monochromatic plane waves of the kind $\exp{\left( i {k_x}  x+i k_z z -i \omega t \right)}$, propagating along the $x$-axis. The dispersion relations, describing the propagation of a $p$-polarized plane wave in the considered system, are  
\begin{equation}
\label{vac}
k_x^2+ k^{(v)2}_z= \frac{\omega^2}{c^2},
\end{equation}
and 
\begin{equation}
\label{sla}
\frac{k_x^2}{\varepsilon_{\perp}(\omega)}+ \frac{k^{(s)2}_z}{\varepsilon_{\parallel}(\omega)}= \frac{\omega^2}{c^2},
\end{equation}
in vacuum and inside the uniaxial crystal, respectively, where $k^{(v)}_z$, $k_{z}^{(s)}$ are the complex wave vectors along the $z$-axis. Imposing the standard interface boundary conditions (the continuity of the electric and the magnetic field parallel to the vacuum-slab interface), we get the relation\cite{Warm},
\begin{equation}
\label{disp}
\epsilon_{\parallel}(\omega) k^{(v)}_z= k_{z}^{(s)}.
\end{equation}
 In the static limit (i.e., $|k_x|\gg \omega/c$), Eq.(\ref{vac}) and Eq.(\ref{sla}) imply 
\begin{equation}
\label{disp}
k_{z}^{(v)} \simeq \pm i k_x, \quad k_{z}^{(s)} \simeq  i \sqrt{\frac{\varepsilon_{\parallel}(\omega)}{\varepsilon_{\perp}(\omega)}} k_x,
\end{equation}
Finally, imposing that the SP amplitudes exponentially decay away from the interface [${\rm Im} k_z^{(v)}>0$ and, ${\rm Im} k_z^{(s)}<0$], and considering Eq. (\ref{disp}), we get 
\begin{equation}
\label{xxxi}
\xi(\omega)=\sqrt{\varepsilon_{\parallel}(\omega)\varepsilon_{\perp}(\omega)}=-1.
\end{equation}
Equation (\ref{xxxi}) represents the existing condition of an SP supported by a uniaxial crystal, and it can be exactly satisfied in the situation where the imaginary part of the dielectric functions are negligible. In this idealistic case, the SPs are excited for $\varepsilon_{\parallel}(\omega)=1/\varepsilon_{\perp}(\omega)$ in the region where $\varepsilon_{\parallel}(\omega)<0$ and $\varepsilon_{\perp}(\omega)<0$. As reported in the main text, the resonances appearing in the electron energy loss spectroscopy (EELS) spectrum, given by the condition ${\rm Re} \xi(\omega)=-1$, are clearly associated with the excitation of SPs.

\section{Ambient stability}

To evaluate the environmental stability of NiTe$_2$, the evolution of atomic force microscopy (AFM) images of mechanically exfoliated flakes [Figs. S6(a)-(d)] was followed on a timescale of up to $10$ days. The AFM results demonstrate that exposure to the atmosphere does not change the morphology of the flakes, as confirmed by the height profile along a specific direction remaining constant with exposure [Fig. S6(e)].
In addition, the EELS spectrum does not display noticeable changes upon exposure to air, congruently with ambient stability inferred by AFM experiments (see Fig. S7). 
\begin{figure}[h]
\centering
\includegraphics[width=0.8\linewidth]{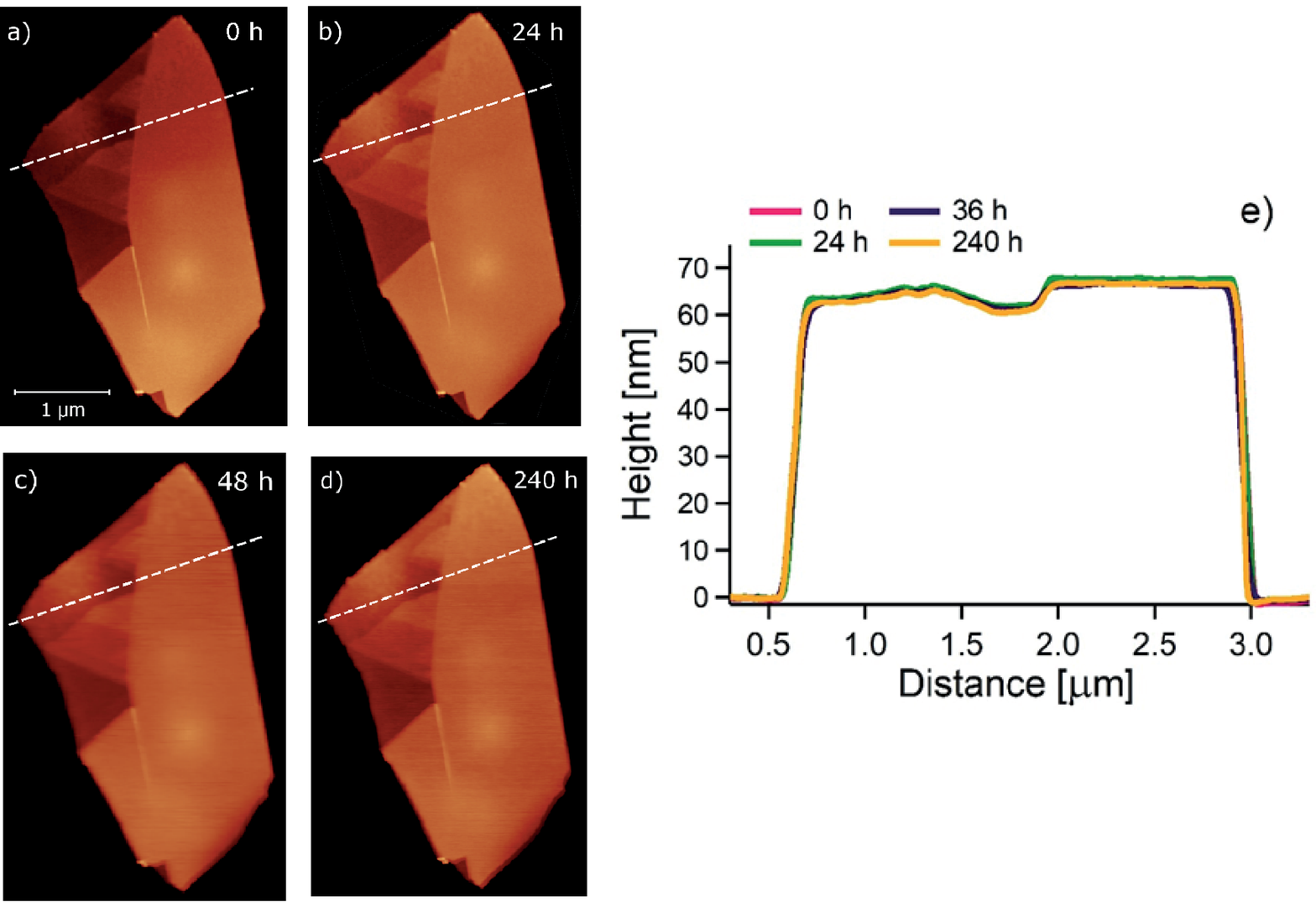}	
\caption{Time evolution of AFM images of a $\sim 60$ nm thick flake of NiTe$_2$. Panel (a) shows the flake immediately after exfoliation, while panels (b)-(d) show the same flake after $24$, $48$ and $240$ hours in air. The dotted white lines indicate the path of the height profile shown in panel (e).}
\end{figure}
\begin{figure}[h]
\centering
\includegraphics[width=0.5\linewidth]{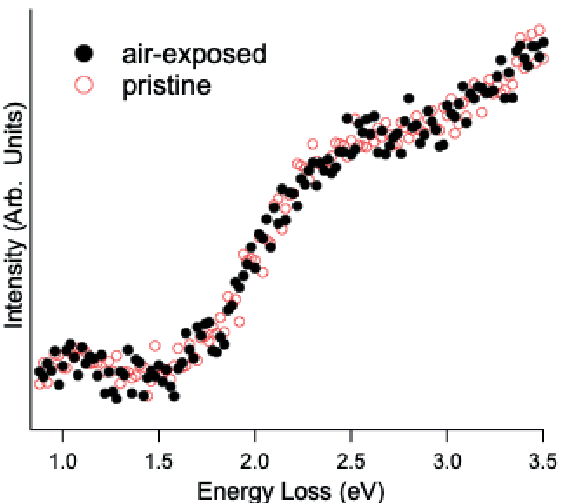}	
\caption{Experimental EELS spectra of NiTe$_2$, at a primary electron beam energy $200$ eV, before and after exposure to air.}
\end{figure}

\section{Connection with interband transitions}
The calculated surface loss spectrum shows three surface plasmon modes at about 0.79 eV, 1.64 eV and 2.22 eV. To trace the origin of these modes, we focus on the imaginary part of the dielectric tensor in the long-wavelength limit (${\bf q} \to 0$). The imaginary part of the dielectric tensor is associated with inter-band transitions, and it can indicate which pair of bands are involved in the transitions corresponding to the surface plasmon energies. 

In the long-wavelength limit, the total dielectric function can be expressed as~\cite{Jia_2020},
\small
\begin{eqnarray}
	\varepsilon_{\parallel/\perp}(\omega) &=& 1+\frac{e^2}{\varepsilon_0 \Omega_{\rm{cell}}}
		\sum_{{m,{\bf q'}}} \left[ 
		\frac{ \frac{\partial f_{m,{\bf q'}}}{\partial E_{m,{\bf q'}}}\left(\frac{\partial E_{m,{\bf q'}}}{\partial k_{i}}\right)^2 }{(\hbar\omega+i\eta_1)^2} \right. \nonumber
		\left. -\hbar^2\sum_{n\ne m}\frac{f_{m,{\bf q'}} -f_{n,{\bf q'}}}{\hbar\omega+E_{m,{\bf q'}}-E_{n,{\bf q'}}+i\eta_2} \frac{ \langle{m,{\bf q'}}|v_{\parallel/\perp}|n,{\bf q'}\rangle\langle{n,{\bf q'}}|v_{\parallel/\perp}|m,{\bf q'}\rangle }{(E_{n,{\bf q'}} -E_{m,{\bf q'}})^2} 
		\right]~. 
		\\ 
		\label{varepsilon_q0}
\end{eqnarray}
Here, $v_{\parallel/\perp}=\frac{1}{\hbar}\frac{\partial {\cal H}_{{\bf q'}} }{\partial k_{\parallel/\perp}}$ denotes the velocity operator. Here $\eta_1$ and $\eta_2$ represents small broadening parameter for intraband and interband contributions, respectively. The imaginary part of the interband dielectric function in Eq.~\eqref{varepsilon_q0} is given by,
\begin{equation}
{\rm Im} \varepsilon_{\parallel/\perp}(\omega) = \sum_{{m,{\bf q'}}} {\rm Im} \varepsilon_{\parallel/\perp}^{(m)}(\omega,{\bf q'}),
\label{varepsilon_inter}
\end{equation}
where, 
\begin{equation}
{\rm Im}\varepsilon_{\parallel/\perp}^{(m)}(\omega,{\bf q'})
= \frac{\pi\hbar^2 e^2}{\varepsilon_{0} \Omega_{\rm{cell}}}\sum_{{m,{\bf q'}}}\sum_{n\ne m}(f_{m,{\bf q'}} -f_{n,{\bf q'}})\frac{ \langle{m,{\bf q'}}|v_{\parallel/\perp}|n,{\bf q'}\rangle\langle{n,{\bf q'}}|v_{\parallel/\perp}|m,{\bf q'}\rangle }{(E_{n,{\bf q'}} -E_{m,{\bf q'}})^2}\delta(\hbar\omega+E_{m,{\bf q'}}-E_{n,{\bf q'}})~.
\label{varepsilon_inter1}
\end{equation}
Using the momentum resolved imaginary part of the dielectric tensor for $\omega$ corresponding to the plasmon frequencies, we can identify the bands and the specific points of the BZ which contribute most to the vertical interband electronic transitions. In Fig. 4 of the main text, we have shown the distribution of ${\rm Im} \varepsilon_{ii}^{(m)}(\omega,{\bf q}=0,{\bf q'})$ projected along the high symmetric direction in the BZ for three plasmon peaks at 0.79 eV, 1.64 eV and 2.22 eV.


\begin{thebibliography} {aa}
\bibitem{HYPE} Hyperbolic metamaterials are extreme anisotropic crystals behaving as a metal along one direction and as a dielectric along the orthogonal direction.
\end{thebibliography}
\end{document}